# Bifurcations Observed in the Spectra of Coupled Electron-Phonon Modes in Multiferroic PrFe$_3$(BO$_3$)$_4$ Subjected to a Magnetic Field

K. N. Boldyrev[1*], T. N. Stanislavchuk[2], A. A. Sirenko[2], D. Kamenskyi[3], L. N. Bezmaternykh[4], and M. N. Popova[1]

[1]*Institute of Spectroscopy, Russian Academy of Sciences, 108840 Troitsk, Moscow, Russia*
[2]*Department of Physics, New Jersey Institute of Technology, 07102 Newark, New Jersey, USA*
[3]*High Field Magnet Laboratory (HFML-EMFL), Radboud University, 6525 ED Nijmegen, Netherlands*
[4]*Kirenskiy Institute of Physics, Siberian Branch of RAS, 660036 Krasnoyarsk, Russia*

We report on bifurcations effect mediated by the electron-phonon coupling in a concentrated rare-earth-containing antiferromagnet, observed in the spectra of coupled 4*f*-electron-phonon modes under the influence of an external magnetic field. The effect was observed in the low-temperature far-infrared (terahertz) reflection spectra of a multiferroic easy-axis antiferromagnet PrFe$_3$(BO$_3$)$_4$ in magnetic fields **B**$_{ext}$||*c*. Both paramagnetic and magnetically ordered phases (including a spin-flop one) were studied in magnetic fields up to 30 T. We show that the field behavior of the coupled modes can be successfully explained and modeled on the base of the equation derived in the frame of the theory of coupled electron-phonon modes, with the same field-independent electron-phonon interaction constant $|W|$ = 14.8 cm$^{-1}$.

A coupling between the electronic system and the lattice vibrations mediates a number of important effects in solids such as classical superconductivity [1], different kinds of the Jahn – Teller effect [2], a splitting of degenerate phonon modes in concentrated transition-metal or rare-earth (RE) compounds under an applied external magnetic field [3-5], delocalization of the electronic states in the energy range of optical phonons and, as a consequence, an observable electronic Davydov splitting [4,6]. Narrow zero-phonon spectral lines in crystals, which are useful probes of a local crystal structure and of different interactions, shift and dramatically broaden with increasing the temperature, mainly because of the electron-phonon coupling [7]. The same coupling is responsible for the crystal-field levels' relaxation and, hence, the lifetimes. The values of these effects and of the multiphonon relaxation rates (also dependent on the electron-phonon interaction) are of primary importance for laser applications, because they influence the gain, output frequency stability, and thermal tunability of a laser [7].

In the case of a resonance between a phonon and an electronic excitation, the electron-phonon interaction results in a formation of coupled electron-phonon modes [5,6,8-11]. An avoided crossing of the electronic crystal-field (CF) level of a rare-earth (RE) ion with the acoustic phonon branch in the middle of the Brillouin zone, which is a signature of a coupled electron-phonon mode, was observed in neutron scattering experiments on TmVO$_4$ [9] and Tb$_2$Ti$_2$O$_7$ [10]. In optical measurements, which probe the Γ point (**k** = 0) of the Brillouin zone, the resonance between an optical phonon and an electronic excitation was achieved by tuning an electronic level by an external magnetic field. In such way, coupled electron-phonon modes were observed in Raman and infrared spectra of a number of RE containing compounds [3-6]. An avoided crossing between a field-dependent Co$^{2+}$ CF excitation level and low-frequency wagging phonons was observed in Raman spectra of Co[N(CN)$_2$]$_2$ in a strong magnetic field of about 20 T [11].

In this Letter, we report on a new effect associated with the electron-phonon interaction, namely, a bifurcation, under an external magnetic field, in the spectrum of coupled electron-phonon modes in a magnetically ordered compound. The effect is observed at terahertz frequencies in the reflection spectrum of a multiferroic antiferromagnetic PrFe$_3$(BO$_3$)$_4$ single crystal. Multiferroic compounds exhibit a mutual interference of the charge, lattice, and spin degrees of freedom and are the most promising candidates for observing new effects mediated by the electron-phonon interaction. In particular, a strong resonance electron-phonon interaction resulting in a coupled phonon – 4*f* electronic excitations has recently been evidenced in multiferroic PrFe$_3$(BO$_3$)$_4$ [12] and TbFe$_3$(BO$_3$)$_4$ [13].

Both mentioned compounds crystallize into the trigonal noncentrosymmetric structure of the natural mineral huntite (space group *R*32), order into an easy-axis antiferromagnetic (AFM) structure (at 32 and 40 K for PrFe$_3$(BO$_3$)$_4$ [14,15] and TbFe$_3$(BO$_3$)$_4$ [16,17], respectively), undergo a spin-flop phase transition at which the antiferromagnetically ordered along the *c* axis Fe$^{3+}$ spins flop onto the *ab* plane (in the magnetic field $B_{SF}$ of 4.5 and 3.5 T for PrFe$_3$(BO$_3$)$_4$ [14] and TbFe$_3$(BO$_3$)$_4$ [16], respectively, at *T* = 4.2 K), incorporate non Kramers RE ions that have crystal-field levels in the energy region of phonons, and belong to a new class of multiferroics [18].

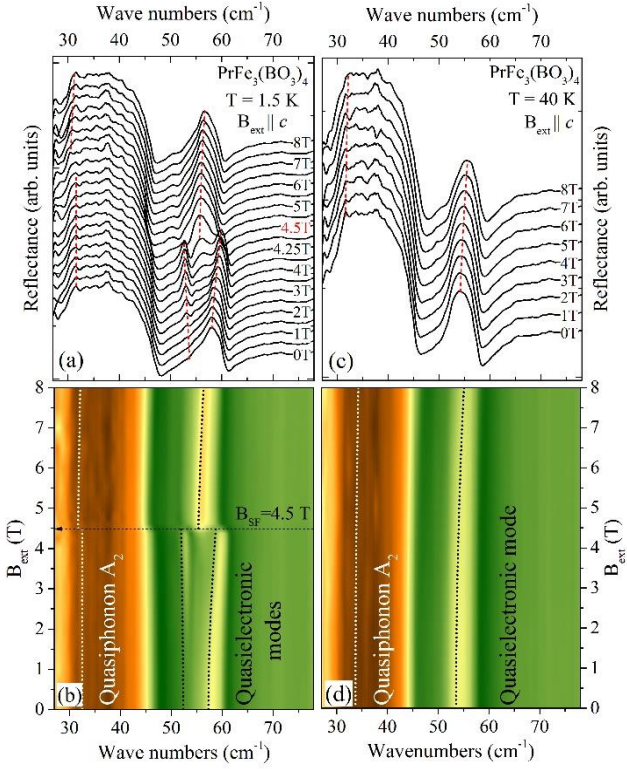

FIG. 1 (color online). The $\pi$-polarized far-infrared reflection spectra of PrFe$_3$(BO$_3$)$_4$ in external magnetic fields $\mathbf{B}_{ext} \parallel c$ up to 8 T and the corresponding reflection intensity maps [dark-brown (dark-green) color corresponds to the reflectance close to unity (zero)] as functions of frequency and magnetic field at (a,b) $T$ = 1.5 K and (c,d) $T$ = 40 K > $T_N$ = 32 K. The spin-flop transition in the field $B_{SF} \approx 4.5$ T is clearly observed in the spectra taken at 1.5 K. Dotted lines represent results of calculations according to Eqs. 1 and 2, with the following set of parameters: (a,b,c,d) $|W|$ = 14.8 cm$^{-1}$, $\omega_{ph}$ = 40 cm$^{-1}$, $E$ = 49 cm$^{-1}$, $g_0 \langle \Gamma_1 | J_z | \Gamma_2 \rangle = 1.9$, and (a, b) $B_{int}$ = 10.5 T for $B_{ext} < B_{SF}$ (for details, see the text).

Effects observed due to the 4$f$ electron-phonon interaction are particularly simple and clear for interpretation in the case of PrFe$_3$(BO$_3$)$_4$ that preserves the R32 ($D_3^7$) structure with one formula unit in the primitive crystal cell down to the lowest temperatures and demonstrates a resonance interaction between nondegenerate electronic and lattice excitations of the same symmetry ($\Gamma_2$ and $A_2$ irreducible representations of the crystal factor group $D_3$, in the notations accepted for electronic excitations and phonons, respectively). In contrast, TbFe$_3$(BO$_3$)$_4$ undergoes a structural phase transition (at ~200 K) into an enantiomorphic space-group pair $P3_121$ ($D_3^4$) and $P3_221$ ($D_3^6$) with three formula units in the primitive cell and exhibits a coupling between a doubly degenerated $E$ phonon and two Davydov multiplets of the $\Gamma_3$ symmetry.

In PrFe$_3$(BO$_3$)$_4$, the singlet ground $\Gamma_2$ and the first excited $\Gamma_1$ (at about 49 cm$^{-1}$) states of the Pr$^{3+}$ ion are well isolated from the other CF states (the next nearest CF level is at 192 cm$^{-1}$ and has the $\Gamma_3$ symmetry [15]), they govern the low-temperature magnetic and magnetoelectric properties of the compound [14,15,19]. The electronic CF excitation corresponding to the $\Gamma_2 \rightarrow \Gamma_1$ optical transition of Pr$^{3+}$ strongly interacts with the lowest-frequency $A_2$ lattice phonon mode associated, mainly, with motions of the heavy Pr$^{3+}$ ions, which is accompanied by a strong mixing of electronic and vibrational wave functions, energy renormalization, and redistribution of intensities in the spectrum (i.e., a coupled electron-phonon mode is formed) [12].

In the present study, we have used the same well-oriented single crystal of PrFe$_3$(BO$_3$)$_4$ as in Ref. [12]. Optical $\pi$-polarized ($\mathbf{k} \perp c$, $\mathbf{E} \parallel c$) reflection measurements in the far-infrared (terahertz) spectral region 20-100 cm$^{-1}$ (0.6 – 3 THz) at the temperatures 1.5 –50 K and external magnetic fields $\mathbf{B}_{ext} \parallel c$ up to 8 T were performed on the U4IR beamline of the National Synchrotron Light Source, Brookhaven National Laboratory, USA, using a Fourier spectrometer Bruker IFS-66v with a liquid helium bolometer (4.2 K) as a detector and an optical Oxford SM4000 superconducting magnet. Unpolarized reflection ($\mathbf{k} \perp c$) over the same spectral interval, at 1.5 K and in strong magnetic fields up to 30 T was studied at the High Field Magnet Laboratory (HFML), the Netherlands. For this study, we used a Fourier-transform spectrometer Bruker IFS-113v combined with a 33-Tesla Bitter electromagnet. The radiation was detected using a silicon bolometer operating at 1.5 K.

Figures 1(a) and 1(c) display representative reflection spectra of PrFe$_3$(BO$_3$)$_4$ in magnetic fields $\mathbf{B}_{ext} \parallel c$ up to 8 T at the temperatures 1.5 K and 40 K > $T_N$ = 32 K, respectively. The corresponding reflection intensity maps in the frequency – external magnetic field axes are presented in Figs. 1(b) and 1(d). In an easy-axis antiferromagnetic state ($T$ = 1.5 K, $B_{ext} < B_{SF}$), iron magnetic moments aligned along the $c$ axis create an internal staggered magnetic field at the praseodymium sites, $\mathbf{B}_{int} \parallel c$, the value of which reaches $B_{int} \approx 10.5$ T at 1.5 K [15]. An external magnetic field $\mathbf{B}_{ext} \parallel c$, being summed with this field, results in effective magnetic fields $B_{eff}^{(i)}$ $B_{eff}^{(1)} = B_{int} + B_{ext}$ at one half of the praseodymium sites (Pr subsystem 1) and $B_{eff}^{(2)} = B_{int} - B_{ext}$ at the other half (Pr subsystem 2), as is shown schematically in Figs. 2(a) and 2(b). At the absence of the electron-phonon interaction, there are two electronic energy branches with energies $\omega_{el,1}$ and $\omega_{el,2}$ corresponding to these two Pr subsystems:

$$\omega_{el,i}^2 = E^2 + 4\mu_B^2 g_0^2 \left|\langle \Gamma_1 | J_Z | \Gamma_2 \rangle\right|^2 \left|B_{eff}^{(i)}\right|^2, \quad i = 1, 2 \quad (1)$$

(here, $E = 49$ cm$^{-1}$ is the energy of the $\Gamma_1$ first excited CF level of Pr$^{3+}$ in paramagnetic PrFe$_3$(BO$_3$)$_4$, $\mu_B$ is the Bohr magneton, $g_0$ is the Lande factor). The energies of these branches converge to the same value $\omega_{el}$ at $B_{ext} \to 0$,

$$\omega_{el}^2 = E^2 + 4\mu_B^2 g_0^2 \left|\langle \Gamma_1 | J_Z | \Gamma_2 \rangle\right|^2 B_{int}^2.$$

As is evident from Figs. 1(a) and 1(b), two electronic branches (observed in the region from 50 to 60 cm$^{-1}$) converge to different wave numbers at $B_{ext} \to 0$, which confirms once more a crucial role of the electron-phonon coupling.

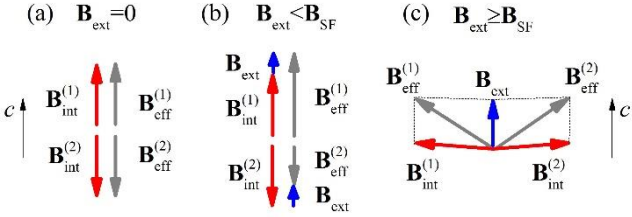

FIG. 2 (color online). Scheme illustrating formation of an effective magnetic field at the two praseodymium subsystems in different regions of the $B$-$T$ phase diagram in the case $\mathbf{B}_{ext} \parallel c$. (a) AFM phase, $B_{ext} = 0$; (b) AFM phase, $B_{ext} < B_{SF}$; (c) spin-flop phase, $B_{ext} \geq B_{SF}$.

To find the frequencies of coupled electron-phonon excitations in the case of nondegenerate phonon mode with unperturbed frequency $\omega_{ph}$ and two electronic excitations having frequencies $\omega_{el,1}$ and $\omega_{el,2}$, one has to solve the following equation obtained on the basis of the results derived in Ref. [8] using the Green's functions method:

$$\left(\omega^2 - \omega_{ph}^2\right) - 2\omega_{ph}|W|^2 \left(\frac{\omega_{el,1}(n_{01} - n_1)}{\omega^2 - \omega_{el,1}^2} + \frac{\omega_{el,2}(n_{02} - n_2)}{\omega^2 - \omega_{el,2}^2}\right) = 0$$

(2)

Here, $W$ is the electron-phonon coupling constant, $n_{0i}$ and $n_i$ are relative populations of the ground and excited states, respectively, of the Pr$^{3+}$ ions in the $i$-th praseodymium subsystem, $i = 1, 2$. In the case of $T = 1.5$ K considered here, $n_1 \approx n_2 \approx 0$ but $n_{01} \approx n_{02} \approx 1/2$ and we rewrite Eq. 2 in the following form:

$$\left(x - x_{ph}\right) = \sqrt{x_{ph}}|W|^2 \left(\frac{\sqrt{x_{el,1}}}{x - x_{el,1}} + \frac{\sqrt{x_{el,2}}}{x - x_{el,2}}\right) \quad (3)$$

where the notations $x_{ph} \equiv \omega_{ph}^2$, $x_{el,i} \equiv \omega_{el,i}^2$ are introduced. Figure 3 presents a graphical solution of Eq. 3. The right part of the equation as a function of $x$ consists of three branches (marked (0), (1), and (2) in Fig. 3), they have singularities at the points $x_{el,1}$ and $x_{el,2}$. These branches are crossed by a straight line representing the left part of Eq. 3, at the three points, $x_0$, $x_1$, and $x_2$, which give the roots of Eq. 3. The branch (2) disappears in the case $x_{el,1} = x_{el,2}$ and a cubic Eq. 3 (or 2) reduces to a quadratic one. There are three situations in our experiment on PrFe$_3$(BO$_3$)$_4$ corresponding to this case, namely, 1) the AFM phase, $B_{ext} = 0$, 2) the paramagnetic phase ($B_{int} = 0$), any $B_{ext}$, and 3) the AFM spin-flop phase, $B_{ext} \geq B_{SF}$.

The first case, which corresponds to the lowest spectral trace in Fig. 1(a), has already been considered in Ref. [12]. Two reflection bands present in a zero magnetic field correspond to the two branches of a coupled 4$f$ electron-phonon mode that is formed below the temperature of ~ 100 K, at which the population of the CF level lying at about 49 cm$^{-1}$ markedly diminishes. As the temperature decreases, the high-frequency quasielectronic branch gains its intensity at the expense of a strong low-frequency quasiphonon branch. It is worth noting that pure 4$f$ electronic excitations are not observable in the reflection spectra, due to their low oscillator strengths (~ 10$^{-6}$ - 10$^{-8}$). The temperature behavior of the spectrum has been successfully modeled using Eq.2 with $\omega_{el,1} = \omega_{el,2} = \omega_{el}(T)$ found from the optical measurements, the Boltzman distribution of populations of electronic levels, $|W| = 14.8$ cm$^{-1}$, and $\omega_{ph}(T) = \omega_{ph}(300\,\text{K}) - 0.019(300 - T)$ with $\omega_{ph}(300\,\text{K}) = 45.2\,\text{cm}^{-1}$ [12].

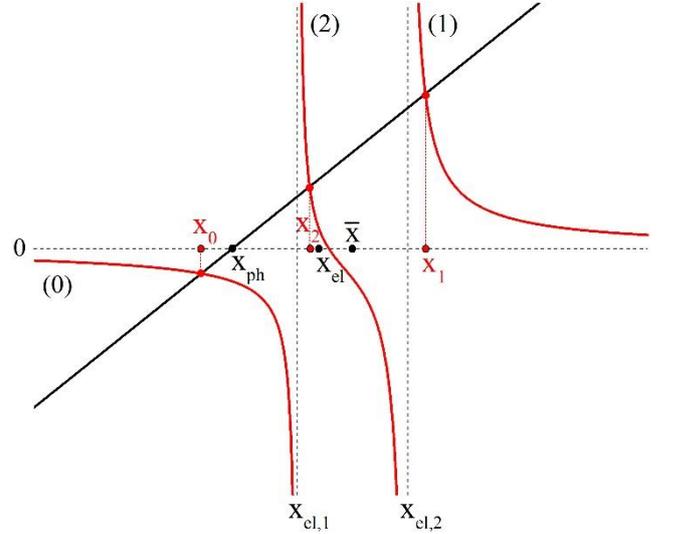

FIG. 3 (color online). Graphical solution of Eq. 3. Roots of the equation are found as abscissas of the points where the straight line crosses branches (0), (1), and (2). The branch (2) disappears in the case $x_{el,1} = x_{el,2}$. $\overline{x} = 1/2(x_{el,1} + x_{el,2})$. For other notations, see the text.

In the second case of a paramagnetic state of PrFe$_3$(BO$_3$)$_4$ ($T = 40$ K, $B_{int} = 0$,), corresponding to Figs. 1(c) and 1(d), an external magnetic field $\mathbf{B}_{ext} \parallel c$ shifts the first excited $\Gamma_1$ CF level to higher frequencies

$$\omega_{el}^2 = E^2 + 4\mu_B^2 g_0^2 \left|\langle \Gamma_1 | J_Z | \Gamma_2 \rangle\right|^2 B_{ext}^2,$$ due to nondiagonal

Zeeman interaction with the ground $\Gamma_2$ level. Correspondingly, a gradual shift of both branches of the electron – phonon mode is observed. We model here this behavior in the frame of the same approach as in the first case, with the same set of parameters, namely, $|W| = 14.8$ cm$^{-1}$, $\omega_{ph}(40\,K) = 40$ cm$^{-1}$, and $E = \omega_{el}$ (40 K, $B_{ext} = 0$) = 49 cm$^{-1}$ for the value of the electron-phonon coupling constant and unperturbed ($|W| = 0$) phonon and electronic excitation frequencies, respectively [dotted lines in Figs. 1(c) and 1(d)].

In the third case of an antiferromagnetic PrFe$_3$(BO$_3$)$_4$ subjected to an external magnetic field $\mathbf{B}_{ext} \parallel c$ with the strength above that of a spin-flop transition, $B_{ext} \geq B_{SF} = 4.5$ T ($T = 1.5$ K), the largest component of an internal magnetic field from the iron subsystem is in the $ab$ plane [see Fig. 2(c)] and does not influence the Pr subsystem (because of zero matrix elements of the operators $J_x$, $J_y$ between the states $\Gamma_1$ and $\Gamma_2$ of Pr$^{3+}$). The $z$ component of the internal magnetic field, as estimates based on the magnetization data of Ref. [14] show, is much smaller than $B_{ext}$ at any $B_{ext}$ in the interval between $B_{SF}$ and 30 T. As a result the behavior of all praseodymium ions in the spin-flop phase is governed by $\mathbf{B}_{ext} \parallel c$. Again, a successful modeling is achieved for the frequencies of the two branches of a coupled electron-phonon mode of PrFe$_3$(BO$_3$)$_4$ ($T = 1.5$ K) in an external magnetic fields parallel to the $c$ axis and having the strength from 4.5 to 22 T (for the high-field behavior, see Fig.4). As the external field grows, the energy gap between the electronic level and the considered lattice phonon mode also grows, the interaction between two excitations weakens, and the quasielectronic mode gains less intensity from the quasiphonon mode, until it disappears in the spectrum at fields above ~ 22 T (see Fig. 4). Weak traces of the quasielectronic mode appear again above ~ 25 T when this mode approaches the lowest-frequency $E$ phonon mode. A weak interaction between these two excitations can take place because of an admixture of wave functions of the $\Gamma_3$ CF level at 192 cm$^{-1}$ to the $\Gamma_1$ wave functions by the $ab$ component of the internal magnetic field at the Pr sites in the spin-flop phase. The magnetic structure of the compound does not change in the range of fields $B_{SF} < B_{ext} < 30$ T that are weaker than the Fe-Fe exchange field $B_{Fe-Fe}$ ~ 100 T [19]. It is worth noting the fact that the same electron-phonon coupling constant $|W| = 14.8$ cm$^{-1}$ is valid throughout a broad range of magnetic fields from zero to 30 T, though a field could, in general, influence matrix elements entering $W$ and, thus, alter the latter. The experiment with the fields up to 30 T enabled us to find a precise value of the matrix element $\langle\Gamma_1|J_z|\Gamma_2\rangle$, namely, the value $g_0\langle\Gamma_1|J_z|\Gamma_2\rangle = 1.9$ gave the best agreement with the experimental data.

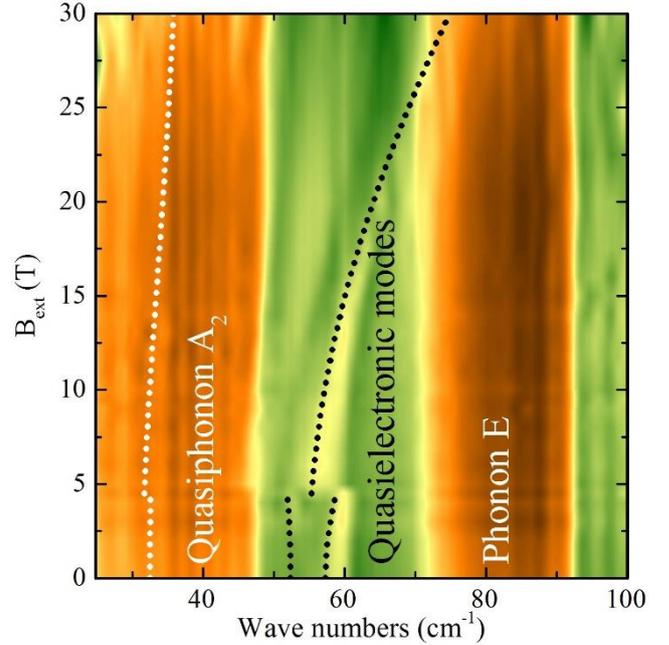

FIG. 4 (color online). Unpolarized far-infrared reflection intensity maps for PrFe$_3$(BO$_3$)$_4$ in external fields $\mathbf{B}_{ext} \parallel c$ up to 30 T, at $T = 1.5$ K. Dotted lines represent results of calculations according to Eqs. 1 and 2, with the following set of parameters: $|W| = 14.8$ cm$^{-1}$, $\omega_{ph} = 40$ cm$^{-1}$, $E = 49$ cm$^{-1}$, $g_0|\langle\Gamma_1|J_z|\Gamma_2\rangle| = 1.9$.

In the case of the electron-phonon coupling in an easy-axis antiferromagnet subjected to an external magnetic field directed along the easy axis, two bifurcation points are present in the spectra of excitations, as can be seen in Figs. 1(a) and 1(b). The first one corresponds to an application of a magnetic field $B_{ext} \neq 0$, which immediately converts a quadratic equation for the coupled electron-phonon modes into a cubic one. A new excitation in the spectrum appears with the frequency $\omega_2^2 \equiv x_2 < x_{el}$. At $B_{ext} \to 0$, $x_2 \to x_{el}$, whereas the root $x_1$ that existed at $B_{ext} = 0$ tends to $x_{el} + O(|W|)$, so that $x_1 > x_{el}$ at any $|W| \neq 0$. Thus, at any $|W| \neq 0$ and $0 < B_{ext} < B_{SF}$ there are two quasielectronic branches in the spectrum of coupled modes in the easy-axis AFM phase of PrFe$_3$(BO$_3$)$_4$, with a gap between them. While the "old" high-frequency quasielectronic branch almost does not change its intensity with growing magnetic field, the "new" low-frequency one gradually gains its intensity. This behavior of intensities demands a special consideration, which is out of scope of the present Letter. The second bifurcation point is at $B_{SF}$. Diminishing of $B_{ext}$ below this value abruptly transforms the quadratic Eq. 2 (with $\omega_{el,1} = \omega_{el,2}$) into the cubic one (Eq. 2, $\omega_{el,1} \neq \omega_{el,2}$) and one quasielectronic excitation splits into two ones with abrupt frequency jumps.

In summary, we have performed a spectroscopic investigation of a multiferroic easy-axis antiferromagnet PrFe$_3$(BO$_3$)$_4$ in magnetic fields $\mathbf{B}_{ext} \parallel c$ up to 30 T, in the far-infrared region that includes frequencies of a praseodymium electronic crystal-field excitation of the $\Gamma_2$ symmetry and the lattice phonon mode of the same symmetry. An interaction between these two excitations results in a formation of a coupled 4*f*-electron-phonon mode. The field behavior of this coupled mode was studied in a paramagnetic (at $T = 40$ K), easy-axis antiferromagnetic ($T = 1.5$ K, $B_{ext} < B_{SF} = 4.5$ T), and a spin-flop ($T = 1.5$ K, $B_{ext} \geq B_{SF}$) phases. We show that in the three cases, namely, (i) the AFM phase in a zero external field, (ii) the paramagnetic phase in any field, and (iii) the AFM spin-flop phase, the field behavior can be successfully modeled on the base of a quadratic equation derived in the frame of the theory of coupled electron-phonon modes, with one field-independent set of parameters. In the case of the easy-axis AFM phase ($T = 1.5$ K, $0 < B_{ext} < B_{SF}$), the quadratic equation converts into a cubic one and bifurcations corresponding to an abrupt appearance of the third root are observed in the terahertz reflection spectra at the two bifurcation points, $B_{ext} = B_{SF}$ and $B_{ext} = +0$. Again, the field behavior of the coupled modes is successfully modeled using the same set of parameters.

*Acknowledgements.* This research was supported by the Russian Scientific Foundation under Grant No 14-12-01033, the President of Russian Federation (MK-3521-2015.2, K.N.B.), and the U.S. Department of Energy under Grant No DE-FG02-07ER46382 (experiments at U4-IR beamline NSLS-BNL, T.N.S. and A.A.S.). The National Synchrotron Light Source is operated as a User Facility for the U.S. Department of Energy under Contract No. DE-AC02-98CH10886. Part of this work was supported by EMFL (contract No. 26211). M.N.P. thanks B.Z. Malkin for helpful discussions.